\documentclass[sigconf]{acmart}

\usepackage{multirow}
\usepackage{float}
\usepackage{array}

\AtBeginDocument{%
  }

\copyrightyear{2025}

\begin{document}

\title{Identifying Video Game Debugging Bottlenecks: An Industry Perspective}

\author{Carlos Pinto Gomez}
\email{carlos-jose.pinto.1@ens.etsmtl.ca}
\orcid{0009-0007-0500-3429}
\affiliation{%
  \institution{École de Technologie Supérieure (ÉTS)}
  \city{Montréal}
  \country{Canada}
}

\author{Fabio Petrillo}
\email{fabio.petrillo@etsmtl.ca}
\orcid{0000-0002-8355-1494}
\affiliation{%
  \institution{École de Technologie Supérieure (ÉTS)}
  \city{Montréal}
  \country{Canada}
}

\renewcommand{\shortauthors}{Pinto Gomez and Petrillo}

\begin{abstract}
    Conventional debugging techniques used in traditional software are similarly used when debugging video games. However, the reality of video games require its own set of unique debugging techniques such as On-Screen Console, Debug Draws, Debug Camera, Cheats and In-Game Menus, and Data Scrubbing. In this article, we provide insights from a video game studio on how 20 seasoned industry game developers debug during the production of a game. Our experiments rely on the recordings of debugging sessions for the most critical bugs categorized as Crashes, Object Behaviors, and Object Persistence. In this paper, we focus on identifying the debugging activities that bottleneck bug resolution. We also identify the debugging tools used to perform debugging techniques. Lastly, we present how different disciplines collaborate during debugging and how technical roles are at the core of debugging. Our thematic analysis has identified game developers spend 36.6\% of their time inspecting game artifacts and 35.1\% of their time reproducing the bug locally.
\end{abstract}





\maketitle

\section{Introduction}
    Video games are developed by integrating the work of multidisciplinary teams, such as game programmers, audio engineers, game designers \citep{gregoryGameEngineArchitecture2018}\citep{karlssonLevelDesignProcesses2023}. During game development, teams are involved in the debugging of various types of bugs varying in categories (e.g., graphical, game logic, and crashes) \citep{trueloveWellFixIt2021a}. Presently, all conventional debugging techniques used in traditional software \citep{decasseReviewAutomatedDebugging1988a} are also used to debug video games \citep{AdvancedDebuggingUnreal2023}\citep{gregoryGameEngineArchitecture2018}. However, there is no research available to understand where the bottlenecks are when game developers debug.

    Currently, debugging research is very limited \citep{vanegueDebuggingVideoGames2023} and focuses on test automation techniques for bug prevention. Furthermore, most of the research focuses on the study of post-mortems from publicly available sources \citep{heinasmakiUsePostMortem2022}\citep{politowskiDatasetVideoGame2020b}\citep{trueloveWellFixIt2021a}. Consequently, this paper presents research involving 20 experienced multidisciplinary game developers. We performed four studies involving the analysis of the studio's bug database, surveys and interviews on the developers, and lastly observations on debugging activities performed. Our aim is to answer the following three research questions:
    \begin{itemize}
        \item \textbf{RQ1} What are the most common bottlenecks when debugging video games?
        \item \textbf{RQ2} How does the use of specific debugging tools impact the debugging efficiency?
        \item \textbf{RQ3} How does cross-disciplinary collaboration affect debugging efficiency?
    \end{itemize}

    Our studies conducted on 20 industry participants have identified that game developers spend 26.4\% of their debugging time inspecting game assets and 10.2\% scripts. In a similar ratio, 25.6\% of their debugging time is spent testing the bug and 9.5\% setting the conditions to reproduce it. The main reason the majority of debugging time is spent on these is to properly identify the system at fault and confirm the role that should fix it. Presently, most first assignments go to the technical team (e.g.: gameplay programmers (36\%), engine programmers (26\%) and technical artists (11\%)) under the conditions that crashes, object behaviors, action and object state bugs are the most prominent. The main bottlenecks encountered during those time-consuming activities were understanding the systems producing bugs, the reproduction of bugs using nondeterministic systems, and the cross-collaboration required due to ownership being spread out.

    In this paper, we find it relevant to first support its content with a brief background on the development of video games. Following, we plan to describe the methods used to retrieve the necessary data to evaluate our three research questions. Specifically, we will present our mixed research approach involving a bug database analysis and three studies on participants such as a survey, observation of debugging sessions, and lastly a semiformal interview. We will then present the results obtained from performing these studies. The following section will be on threats to validity since these results pertain to a specific studio and population. Lastly, we will discuss these industry observations and the conclusions that can be drawn from them.

\section{Background}
    Common knowledge is that video games vary in a combination of one or many genres such as Action, Adventure, Role-Playing, Simulation, Strategy and Shooter \citep{arsenaultVideoGameGenre2009}\cite{apperleyGenreGameStudies2006}\citep{gregoryGameEngineArchitecture2018}. However, this difference is mainly due to the game artifacts which produce unique visual aesthetics, interactivity demands from players, game rules, and systems \citep{apperleyGenreGameStudies2006}.

    The game artifacts used in game development vary from game to game according to \citet{mcdonaldDescribingOrganizingMaintaining2021}. These can be categorized into Game Assets, Front-End, Scripts and Gameplay Code, Artificial Intelligence (AI) as stated by \citet{gregoryGameEngineArchitecture2018}. More specifically, Game Assets are the 2D/3D models, skeleton hierarchy (rig), sounds, visual effects, animations, narrative scripts, game world, and game objects. Front-End is the graphical user interface (GUI), the head-up display (HUD), and the menus. Scripts and Gameplay Code are the instructions to set game design rules, scripted events in the level, game flow, and subsystems' logic. Artificial intelligence (AI), automates game logic for non-playable characters.

    Currently, traditional debugging techniques span program logging, assertions, breakpoints, profiling, and heap analysis \citep{decasseReviewAutomatedDebugging1988a}. Similarly, these conventional debugging techniques are also used when debugging video games \citep{AdvancedDebuggingUnreal2023}\citep{gregoryGameEngineArchitecture2018}. However, video games require an additional set of techniques (e.g.: on-screen console, debug draws, debug camera, cheats and data scrubbing) as described by \citet{gregoryGameEngineArchitecture2018}. Research into advanced debugging techniques for traditional software has led to advances in fault localizations techniques \citep{wongSurveySoftwareFault2016} but video game debugging has still lengths to cover \citep{vanegueDebuggingVideoGames2023}.
        
\section{Methodology}

    To answer the three research questions, four studies were constructed. The first focused on extracting the bugs from a database management system to understand the top categories of bugs and who they are assigned to. The second study surveyed participants about their experience using the respective game engine and their professional experience. The third study focused on observing recordings from debugging sessions and thematically analyzing them to extract conclusions. The fourth study was about conducting interviews to get feedback on how roles collaborate with each other during those sessions.
    
    \subsection{Study 1: Bug Database Extraction}
        The participating studio's bug management system had 921 bugs based on a snapshot taken on June 11, 2025 and can be seen on Table \ref{tab:bug-count} on page \pageref{tab:bug-count}. Bugs varied in status, with some being new, others being worked on, and finally some closed. These bugs were categorized according to the taxonomy of 20 categories of bugs provided by \citet{trueloveWellFixIt2021a}. Bugs were logged by the Quality Assurance (QA) team assigned to the project. All bugs contained fields such as description, summary, assignee, severity, and others important to help the studio prioritize efforts. Throughout the validation of the project, two new categories emerged: Object Behavior and World Traversal. Object Behavior is when two or more objects do not behave as intended when interacting. World Traversal is when regions of the map that should be reachable are not due to world navigation limitations. For severity, a low severity is considered an isolated divergence from the intended system behavior. Medium severity is considered to be a divergence from intended behavior that affects multiple systems. High is considered to have a significant impact on the player's ability to play the game caused by issues in multiple core systems. Critical severity is when the player cannot play the game. Lastly, the assignment of bugs was done by the QA team with the suspicion of who might be able to fix the bug.

        \begin{table}[h]
            \centering
            \begin{tabular}{lcccccc}
                \hline
                \textbf{Bug Category} & \textbf{Crit.} & \textbf{High} & \textbf{Low} & \textbf{Med.} & \textbf{Total} \\
                \hline
                Crash & 127 & 45 & 2 & 21 & 195 \\
                Object Behavior & 5 & 40 & 25 & 89 & 159 \\
                Action & 3 & 39 & 4 & 39 & 85 \\
                Object State & 3 & 23 & 12 & 35 & 73 \\
                UI & 1 & 14 & 8 & 23 & 46 \\
                Camera & 1 & 3 & 12 & 21 & 37 \\
                Object Pos. & 0 & 9 & 10 & 14 & 33 \\
                Audio & 0 & 11 & 6 & 15 & 32 \\
                Collision & 1 & 10 & 7 & 14 & 32 \\
                Graphics & 6 & 7 & 7 & 12 & 32 \\
                Object Persist. & 1 & 18 & 1 & 11 & 31 \\
                Performance & 4 & 10 & 8 & 7 & 29 \\
                World Traversal & 8 & 14 & 1 & 5 & 28 \\
                Event Trigger & 0 & 15 & 2 & 7 & 24 \\
                AI & 0 & 2 & 2 & 17 & 21 \\
                Compatibility & 1 & 12 & 1 & 4 & 18 \\
                World Value & 0 & 4 & 1 & 7 & 12 \\
                Exploit & 2 & 3 & 0 & 5 & 10 \\
                World Info & 0 & 3 & 2 & 3 & 8 \\
                OOB & 0 & 5 & 1 & 1 & 7 \\
                Interrupt & 1 & 2 & 1 & 2 & 6 \\
                Event Occurrence & 0 & 1 & 1 & 1 & 3 \\
                \hline
                Total & 164 & 290 & 114 & 353 & 921 \\
                \hline
            \end{tabular}
            \caption{Bug Categories and Their Severity Counts}
            \Description{A table with Bug Categories and respective occurrences count by severity. The last column is the Total count.}
            \label{tab:bug-count}
        \end{table}
        
    \subsection{Study 2: Survey}
        This research surveyed 20 game developers who actively created artifacts to be integrated in the game. This narrows down the participation to seven roles having a technical background or those with the ability to submit changes to the engine through tools. The survey consisted of three questions with anonymity being respected. The questions focused on asking about their role, years of experience in the role, and years of experience in using Unreal Engine. Of the 21 surveys sent, 20 participants agreed to share their answers, while 1 refused to share their answers. The survey was sent internally by sharing the survey link and 2 months were given to collect responses. 
        
        As seen in Figure \ref{tab:role-experience} on page \pageref{tab:role-experience}, the majority of our participants have between 4 to over 10 years of experience in their current role and as such their preferred debugging techniques may vary based on the games they've debugged in the past. On the other hand, they have about 1-6 years of experience using the Unreal Engine to develop games. As such, the roles studied have a solid understanding of how to develop games but may be new to some of the debugging tools available in the Unreal Engine editor.

        \begin{table}[h]
            \centering
            \begin{tabular}{lccc}
                \hline
                \textbf{ID} & \textbf{Role} & \textbf{Yrs. Exp} & \textbf{Yrs. UE Exp.} \\\hline
                P1 & Tech. Level Designer & 10+ & 1-3 \\\hline
                P2 & Tech. Game Designer & 7-10 & <1 \\
                P3 & Tech. Game Designer & 10+ & 1-3 \\
                P4 & Tech. Game Designer & 4-6 & 1-3 \\\hline
                P5 & Tech. Artist & 10+ & 1-3 \\
                P6 & Tech. Artist & 4-6 & <1 \\
                P7 & Tech. Artist & 10+ & 1-3 \\\hline
                P8 & Level Designer & 10+ & 1-3 \\
                P9 & Level Designer & 10+ & 4-6 \\\hline
                P10 & Gameplay Prog. & 10+ & 1-3 \\
                P11 & Gameplay Prog. & 1-3 & 1-3 \\
                P12 & Gameplay Prog. & 1-3 & 4-6 \\\hline
                P13 & Game Designer & 10+ & 1-3 \\
                P14 & Game Designer & 7-10 & 1-3 \\\hline
                P15 & Engine Prog. & 10+ & 4-6 \\
                P16 & Engine Prog. & 10+ & 4-6 \\\hline
                P17 & Art Designer & 1-3 & 4-6 \\
                P18 & Art Designer & <1 & 7-10 \\
                P19 & Art Designer & 10+ & 1-3 \\
                P20 & Art Designer & 4-6 & 7-10 \\\hline
            \end{tabular}
            \caption{Participant Information}
            \Description{A table withe the list of Participants from this organized by ID, Role, Years of Experience in Role and Years of Experience using the Unreal Engine.}
            \label{tab:role-experience}
        \end{table}
    
    \subsection{Study 3: Observations}
        The observational study involved recording the screen of the game developer while debugging. This study focused on the most frequent and severe bugs encountered during development, since those are the ones being debugged actively and had the highest chance of being fixed during the time provided for this study. As such, not at all participants were involved in the debugging of those categories. In this case, only a total of 4 participants (from the subset of 20) participated in having their screens recorded while they debugged. 3 of those participants recorded their screen using the Open Broadcaster Software (OBS) solution and 1 participant requested that their screen be recorded through a video call while sharing the screen. Recordings allowed for microphone usage, but no webcam was required as the focus was to analyze the screen, and if needed, their real-time voice-overs sometimes provided extra details of information. A debugging recording session started when the developer began reading the bug and ended either when interrupted or when they completed their debugging session. These sessions ended when either the game developer identified a fix or when they required another role to fix it.

        The recordings were obtained in a locally playable format such as MP4 and were stored on the employer's computer for safety. These were replayed and analyzed three times in order to first understand the themes, secondly map activities, and thirdly to finalize the theme analysis. The activities were first coded in a spreadsheet by entering the activity code of the debugging event based on the coding book in Table \ref{tab:activity-coding} on page \pageref{tab:activity-coding}. Following, we added for each activity the start time, the end time, the tool used, the debugging technique used, the details of the activity and the reason for moving on to the next one. The total amount of recorded episodes was 32 and for a total of 188.7 minutes focusing on Crashes, Object Behaviors, Object State, and UI.

        \begin{table}[h]
            \centering
            \begin{tabular}{p{1.65cm}p{2.85cm}p{2.85cm}}
                \hline
                \textbf{Activities} & \textbf{Start Criteria} & \textbf{End Criteria} \\\hline
                 Testing & Reading a the bug & Finish reproducing the bug \\\hline
                 Collaboration & Begin chatting with a peer & End conversation \\\hline
                 Inspection & Open the game system & Closing the viewer \\\hline
                 Consult & Opening the consulting medium & Closing the information resource \\\hline
                 Edit & Open the window to edit & Leaving the edit window \\\hline
                 Misc & Starting an unrelated activity & Ending the unrelated activity \\\hline
            \end{tabular}
            \caption{Coding Book}
            \Description{A table describing the Coding Book used to thematically organize the activities observed during the Debugging Sessions.}
            \label{tab:activity-coding}
        \end{table}

        After coding the activities from the debugging sessions, we proceeded to categorized them under themes to better categorize the objectives game developers are trying to meet. The first goal developers had was trying to better understand the fault, the impact on systems, and the conditions that trigger the bug. The second goal was to locate the system that is causing the bug. The third goal was to fix the fault that caused the bug. Throughout debugging, game developers could be interrupted or begin working with others on non-debugging tasks.
        
    \subsection{Study 4: Post Interviews}
        The final study involved interviewing at least one participant from each of the seven roles identified previously. For this study, 10 participants (a subset of the 20) wished to be interviewed about the bottlenecks they faced when debugging, what tool shortcomings they saw, and how they collaborated. The interviews were semi-structured by asking a question and letting participants answer based on past experience or a recent debugging episode. On average, the interviews lasted 12 minutes.

\section{Results}
    \subsection{Bottlenecks in Video Game Debugging}
        After analyzing the observations, we identified two main bottlenecks during debugging: Inspection activities totaling 41.7\% and testing activities totaling 35\% of the total debugging time. Specifically, the two most time-consuming activities were inspecting components (e.g., game assets), which took 26.4\% of total time, and testing the bug, which took 25.6\% of total debugging time. Figure \ref{tab:activity-time-distribution} on page \pageref{tab:activity-time-distribution} identifies that collaboration (9\% of total time), misc (5.4\% of total time), edit (3.7\% of total time) and consult (2.3\% of total time) were not major bottlenecks when debugging.
        
        \begin{table*}[t]
            \centering
            \begin{tabular}{lcc|lcc}
            \hline
            \textbf{Activity} & \textbf{Total Time (min)} & \textbf{\% of Total} & \textbf{Sub-Activity} & \textbf{Total Time (min)} & \textbf{\% of Total} \\ \hline
            \multirow{6}{*}{Inspection} 
                & \multirow{6}{*}{84.1} & \multirow{6}{*}{41.7} & component\_inspection & 49.7 & 26.4\\ 
                &                      &                       & script\_inspection    & 19.2 & 10.2 \\ 
                &                      &                       & performance\_inspection & 8.2 & 4.4 \\ 
                &                      &                       & level\_inspection     & 5.5  & 2.9 \\ 
                &                      &                       & performance\_tracing  & 1.0  & 0.5 \\ 
                &                      &                       & log\_inspection       & 0.5  & 0.3 \\ \hline
            \multirow{2}{*}{Testing} 
                & \multirow{2}{*}{66.1} & \multirow{2}{*}{35} & test\_bug            & 48.3 & 25.6 \\ 
                &                      &                       & setup\_bug           & 17.8 & 9.5  \\ \hline
            \multirow{2}{*}{Collaboration} 
                & \multirow{2}{*}{16.9} & \multirow{2}{*}{9.0}  & chat\_consult        & 11.9 & 6.3 \\ 
                &                      &                       & post\_consult        & 5.0  & 2.7 \\ \hline
            Misc & 10.2 & 5.4 & misc                & 10.2 & 5.4 \\ \hline
            \multirow{2}{*}{Edit} 
                & \multirow{2}{*}{6.0}  & \multirow{2}{*}{3.7}  & component\_edit      & 4.0  & 2.2 \\
                &                      &                       & script\_edit         & 2.0  & 1.6 \\ \hline
            \multirow{2}{*}{Consult} 
                & \multirow{2}{*}{4.2}  & \multirow{2}{*}{2.3}  & bug\_consult         & 2.8  & 1.5 \\
                &                      &                       & doc\_consult         & 1.4  & 0.8 \\ \hline
            \textbf{Total} & \textbf{188.7} & \textbf{100.0} & -- & \textbf{188.7} & \textbf{100.0} \\ \hline
            \end{tabular}
            \caption{Time and Percentage Distribution Across Activities and Sub-Activities}
            \Description{A table describing the time the participants spent in each type of activities. On the left half, you have the activity types with its total time and percentage. On the right half you have the sub-activities with the total time and percentage of total as well.}
            \label{tab:activity-time-distribution}
        \end{table*}

\begin{figure*}[t]
            \centering
            \includegraphics[width=1\linewidth]{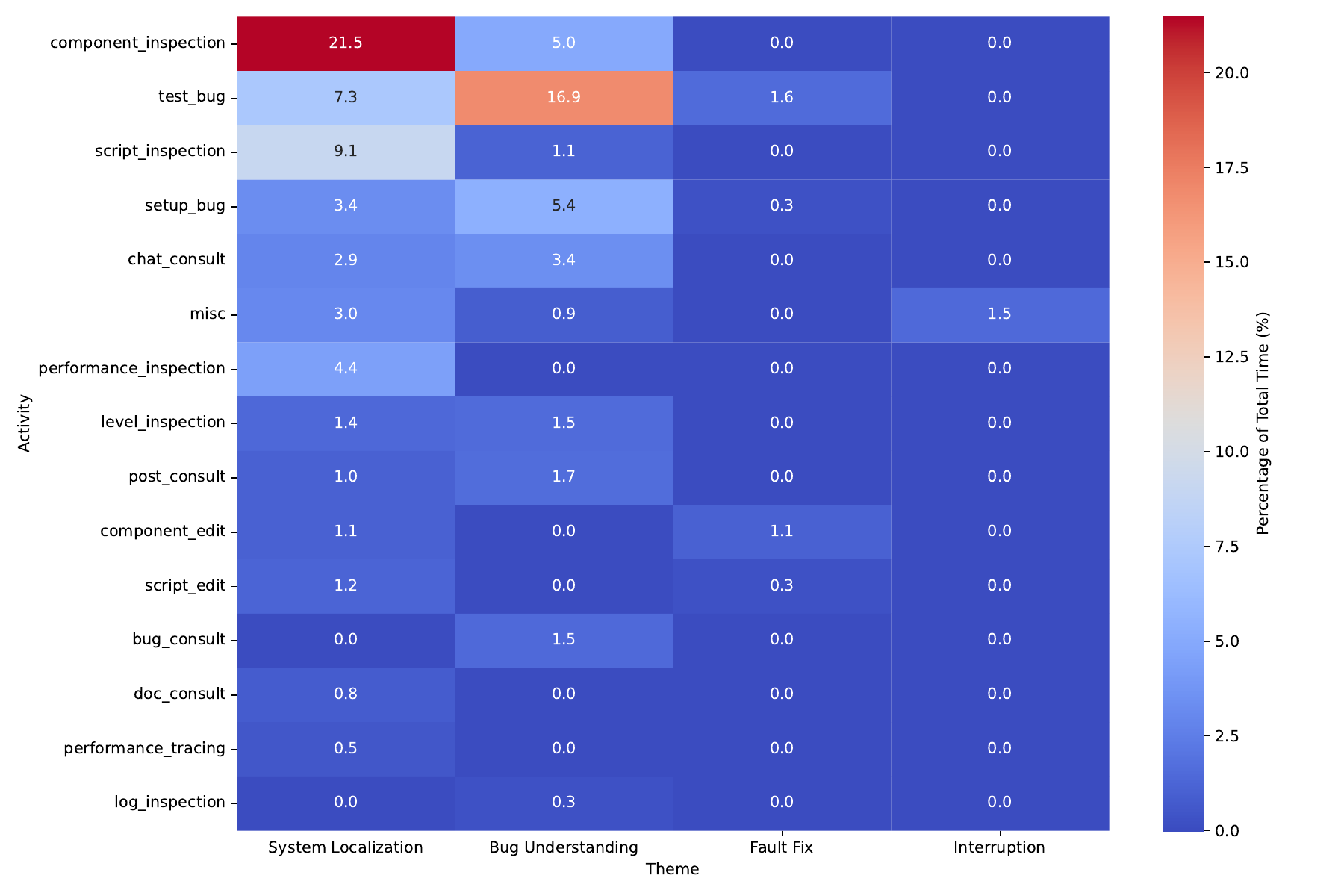}
            \caption{Elapsed Time (\%) Spent on Theme by Activity}
            \Description{A heatmap with the yellow and green colors identifying the activities (y-axis) with the most time spent in relation to the themes (x-axis) identified in the coding book.}
            \label{fig:heatmap-theme-activity-minutes}
            \vspace{10pt}
        \end{figure*}

        The most time-consuming theme was system localization, taking a total of 108 minutes from the total of 188.7 minutes of debugging time. The second most time-consuming theme was bug understanding, which took 70 minutes. The third was the time taken to fix the fault in 6 minutes. Lastly, 2 minutes of interruptions. In terms of which activities were bottlenecks to these goals, we can see in Figure \ref{fig:heatmap-theme-activity-minutes} on page \pageref{fig:heatmap-theme-activity-minutes}, inspecting components was a common activity (40 min) when trying to localize the system at fault shared with trying to understand a bug. Inversely, testing a bug was more time consuming when trying to understand a bug than when trying to localize the system at fault. We can safely say inspection activities bottleneck system localization but testing ones bottleneck bug understanding.

        \begin{figure*}[t]
            \centering
            \includegraphics[width=1\linewidth]{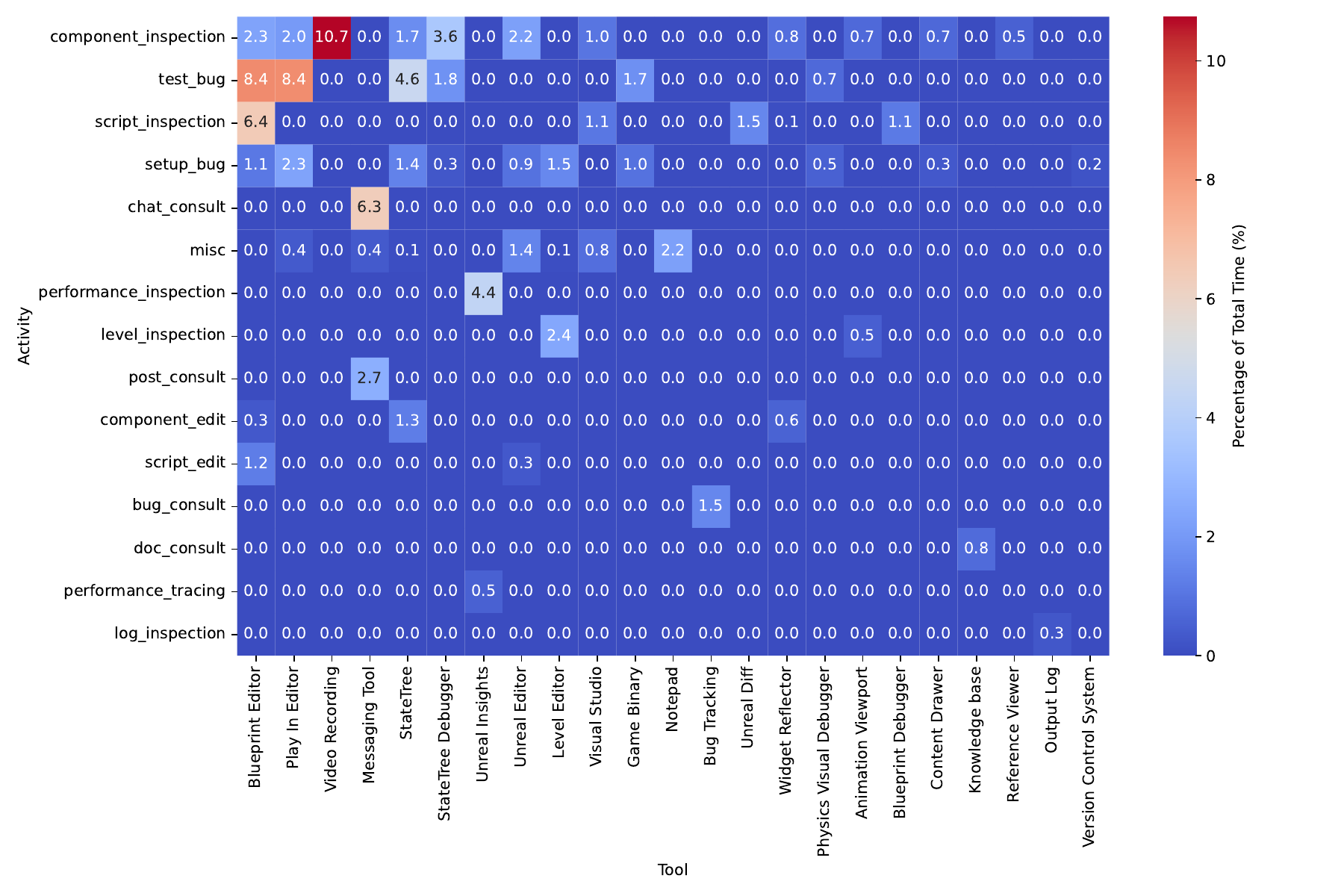}
            \caption{Elapsed Time (\%) Spent on Tool by Activity}
            \Description{A heatmap with the yellow and green colors identifying the activities (y-axis) with the most time spent in relation to the tools being used (x-axis).}
            \label{fig:heatmap-tool-activity-time}
        \end{figure*}

        \begin{figure*}[t]
            \centering
            \includegraphics[width=1\linewidth]{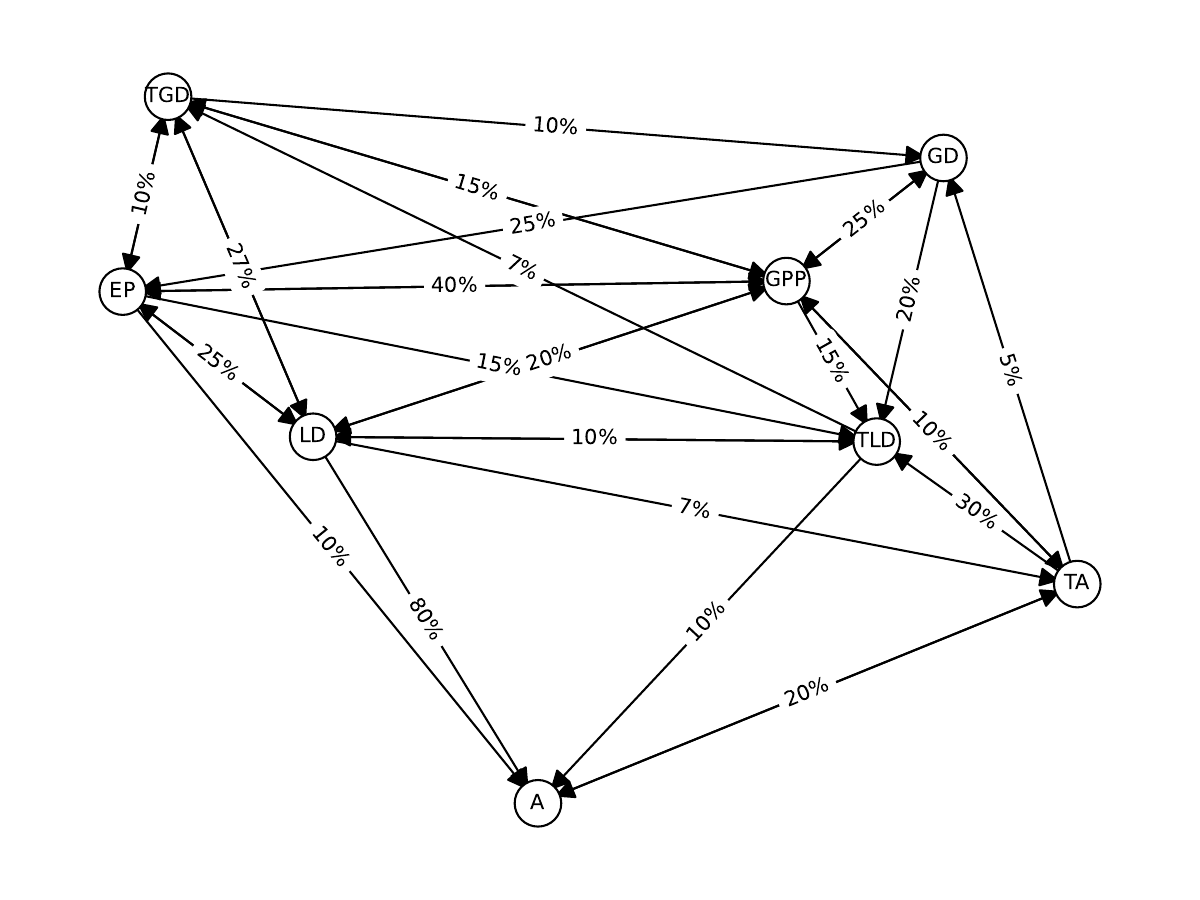}
            \caption{Cross-Collaboration Frequency Ratios}
            \Description{Directed Graph with the ratio of collaboration between roles from our study.}
            \label{fig:collaboration-graph}
        \end{figure*}

    \subsection{Impact of Debugging Tools and Data}
        Game developers used the tools available in the game engine to debug. These tools indirectly used debugging techniques, such as debug draws, to better represent the game data being scrutinized. Figure \ref{fig:heatmap-tool-activity-time} on page \pageref{fig:heatmap-tool-activity-time} exposes how component\_inspection and script\_inspection activities were most often debugged through Editor tools, Content Viewers, and Video Recordings. Specifically, for component-\_inspection, doing playback on the debugging video recording took 20 min of debugging time and allowed analyzing the game state. On the other hand, the Blueprint editor and locally running the game through the editor were used most often to test the bug and less to inspect components during debugging. 
        
        The reality is that not all roles use the same tools to debug. Technical Artists debugged using state trees to debug animation and PIX or Unreal Insights to debug rendering issues. Technical Level Designers used data directly from the world loaded in the editor to understand objects. Technical Game Designers used the Blueprint Debugger and Unreal Editor to debug logic bugs. However, when they debugged performance issues, they used Unreal Insights to analyze CPU/GPU frames. Game Designers debugged the data used to configure the gameplay logic and relied on tools that provide the history of data changes. Level Designers had to debug bugs in the game binaries and in the game running on the editor, but did not express the need for any specific debug tool. Artists were often at a disadvantage when faced with the complexity of debugging tools such as PIX to debug GPU frames due to their counterintuitive UI and relied on technical roles to debug. Engine Programmers used an Integrated Development Environment (IDE) to run the source code of the engine and then debugged as we would traditional software. Gameplay Programmers were sometimes required to interact with external systems used to simulate gameplay logic and consequently used the provided third-party solutions to debug them. a common theme, was the vast amount of data to analyze often leads to game developers requesting for a 'digest tool'.
        
    \subsection{Influence of Cross-Disciplinary Collaboration}
        From the 10 game developers who were interviewed, we gathered that technical roles were often those involved in debugging. Technical Artists (TAs) fixed issues 50-70\% independently, handling data problems while delegating 30\% of code issues due to skill gaps. Technical Level Designers (TLDs) had about 50\% autonomy. Technical Game Designers (TGDs) exhibited 80\% autonomy, delegating 20\% after confirming someone else (typically a Programmer) could fix the fault. Game Designers (GDs) were 80\% of the time autonomous during the early phases of the project due to Blueprint interactions, dropping to 60\% in later stages requiring the help of the programmer to complete the logic implementation. Level Designers (LDs) fixed 40\% of issues, while Artists (A) resolved almost 100\% of their bugs independently when it came to Art asset changes in a DCC tool. Gameplay Programmers (GPPs) maintained 90\% autonomy, delegating 10\% of data issues that game designers used to configure systems. Engine Programmers (EPs) often support other teams that need engine work and are 100\% autonomous when they need to fix their own systems. 

        Figure \ref{fig:collaboration-graph} on page \pageref{fig:collaboration-graph} displays the collaboration ratios of the roles when debugging. By analyzing the debugging sessions and interviews, Gameplay Programmers, Engine Programmers and Level Designers were at the core of collaboration requests. Game Designers engaged frequently with Gameplay Programmers (e.g. approximately 20-25\% of their collaborations) to resolve gameplay logic bugs. Meanwhile, Level Designers worked with Technical Level Designers (e.g. around 10\%) for level-related issues. Technical Artists often collaborated with Artists, (e.g., 20\%) to support asset integration and rendering matters. Technical Artists also collaborated with Gameplay Programmers (e.g. 10\%), when the gameplay logic was linked to their asset integration. Engine Programmers directly collaborated with Gameplay Programmers due to the Engine Programmers creating the underlying systems being used by the gameplay logic.
        
\section{Threats to Validity}
    Our research project is based on the analysis of a video game in development. As such, the focus is mostly on the gameplay aspect and relies significantly on game designers, gameplay programmers, engine programmers, and technical designers. Consequently, other roles such as technical artists, level designers, animators, and sound engineers were not studied proportionally to the previous roles. Consequently, there is a bias on the types of debugging sessions that were observed which focused on object related bugs, crashes, and HUD issues. In subsequent phases, the focus will be on debugging performance degradations, level inconsistencies, ai, etc.
    
    The second threat is that, while findings are identified in a specific genre and game engine, learning can still be abstracted to bug categories and game development principles. However, the tools used are often replicated in other game engines, but may not support the same debugging techniques. For example, some studios use their own custom debugging solutions while others use third-party ones to debug, say physics mechanics.

    Lastly, this research involves participants who are veterans in the industry and currently work on the same project within the same studio. As such, the findings may be difficult to generalize due to the consequent small population. Nevertheless, in this paper, the evaluation of the debugging sessions through our coding book is applicable to other video game debugging sessions notwithstanding the studio or game genre. Furthermore, the fact that the Misc, Edit, and Consult activity categories are small in proportion to other categories means that a more diverse population is needed.

\section{Discussions}
    The results show that bottlenecks occur during the fault localization phase of the debugging. Specifically, reproducing bugs and inspecting components are the most time-consuming activities for debugging. This arises from the complexity of the system interactions to create a gameplay experience, and faults may be located in a system while multiple conditions are at play. As such, the main difficulty in localizing faults is in reproducing the conditions in all systems.
    
    Developers that debug are often doing so from the Engine editor's tools. In some cases, they integrate their editor with external debugging tools by attaching it to the game process. These tools require technical knowledge and can bottleneck the debugging of bugs if experience is lacking. Likewise, the data extracted during these debugging sessions are currently not structured to facilitate debugging. It is not uncommon for developers to extend the engine to support pain points faced during debugging. This could lead to "rubberband" solutions in which developers may talk to the data to help them better understand the systems. Furthermore, these findings could also be extended to other game engines that expose their systems through an editor such as \href{https://unity.com}{Unity} and \href{https://godotengine.org}{Godot}.
    
    Since collaboration is at the heart of game development, it is also at the heart of debugging. Focusing on bug reproduction, developers can share the conditions used to trigger the bugs. However, the interviews also hinted at a potential bottleneck in initiating collaboration, identifying the "right person" and effectively communicating the complex context of the bug. This suggests that while collaboration is effective, there is room for improving the mechanisms that facilitate it, such as better internal documentation, knowledge-sharing platforms, or even dedicated swarm debugging sessions \citep{Petrillo2019} for particularly challenging cross-disciplinary issues.
    
    As such, storing game state data from the systems could increase the collaboration of roles when debugging. Similarly, it can validate the fixes that they implement. These data can be seen as a time lapse of states that can be recorded by testers and shared between game developers or even enhanced using debugging tools to inspect layers of these states.

\section{Related Work}
    Compared to \citet{alaboudiWhatConstitutesDebugging2023}, the bottlenecks that affect most traditional software and video games are the debugging time spent on testing activities with 20\% and 35\%, respectively. \citet{politowskiGameIndustryProblems2021} studied 200 postmortems, uncovering that 46\% of the problems faced during game development lies in the use of tools. Many authors have investigated how to improve the tooling for game development. For example, \citet{machadoEvaluationRecommenderSystem2019} set out to improve game design prototyping by using an AI that recommends existing game mechanics. \citet{lucas2017stay} proposed 3Buddy, a co-creative level design tool aimed at assisting game developers in generating innovative and engaging levels. \citet{engstromHaveDifferentKind2019} proposed Deig Writing Companion that enables script iterations and writing dialogues based on play-through traces, similar to a writing a screenplay. In addition to that, there are external tools that already support game development, such as Wwise for audio \citep{wangSoundDesignVideo2018} and Maya and 3DS Max for animation \citep{cooperGameAnimVideo2021}. However, according to the latest study on debugging tools, \citet{vanegueDebuggingVideoGames2023} did not identify any tools that support bug reproduction activities.

\section{Conclusion}
    Debugging video games is a multidisciplinary activity that requires technical and deep knowledge about how systems interconnect. Our findings indicate that the roles most involved in debugging are technical in nature due to the tools needed to identify the systems causing the issues. Roles face bottlenecks when reproducing the conditions in the systems that trigger faulty behaviors. Any future work on improving developers' ability to debug should consider that the engine tools are primordial to debugging games today. More specifically, system localization being the main pain point in debugging, developers rely on inspection tools and runtime execution of the game as their means to attempt to replicate conditions that trigger bugs. Developers collaborate with each other by sharing screenshots, coordinates, and steps to reproduce. Most often testers are at the forefront of bug identification and could be leveraged to record the game states to facilitate bug reproduction. However, the findings in this paper are related to a particular project, but can be generalized to other projects with similar bug categories.
    
    The subsequent work on this topic will focus on tackling the bug reproduction bottleneck in the testing and setting up bugs in a local context. We believe game developers are well supported in the tools they must use when debugging a certain bug category. However, system localization having bug reproduction and inspection bottlenecks, our focus will be on extracting the game state and replay conditions to cut down on the time developers will take in reproducing bugs.

\bibliographystyle{ACM-Reference-Format}
\bibliography{references}

\end{document}